\documentclass[aps]{revtex4}
\usepackage{amsfonts}
\usepackage{amssymb}
\usepackage{amsmath}
\usepackage{graphicx}
\usepackage{epsfig,amsmath}
\usepackage[bookmarksnumbered,linktocpage,pdfstartview=FitH]{hyperref}

\begin{document}

\title{On the cavity evolution and the Rayleigh--Plesset
equation in superfluid helium}
\author{Sergey K. Nemirovskii\thanks{%
email address: nemir@itp.nsc.ru}}
\affiliation{Institute of Thermophysics, Lavrentyev ave, 1, 630090, Novosibirsk, Russia,
Novosibirsk State University, Novosibirsk,\\
National Research University MPEI, Moscow, Russia.}
\date{\today }

\begin{abstract}
On the basis of the two-fluid hydrodynamics, an analogue of the famous
Rayleigh-Plesse equation for the dynamics of a spherical bubble in
superfluid helium is obtained. The mass flow velocity $v$ and the velocity
of the normal component $v_{n}$ were chosen as independent variables. Due to
the two-fluid nature of HeII, the cross terms in the evolution equation
for the boundary position $\ R(t)$ appeared, which were absent
in classical Rayleigh-Plesset equation in ordinary fluids. One of them
renormilizes the coefficient in front of $(dR/dt)^{2}$. Another
additional term formally coinciding with the viscous term, describes
the attenuation of the boundary oscillations. This "extra-damping" term, greatly
exceeding the usual viscous term, leads to a significant difference in the
dynamics of cavity compared to HeI. In particular, this results in the
interesting effect of abnormal suppression of oscillations of the
vapor--liquid boundary observed in many works. There is also an additional
term proportional to the squared velocity of the normal component, which is
independent of the derivative $dR/dt$, and can be included in the pressure
drop. Its physical meaning is that it describes a "Bernoulli" -like pressure
created by the flow of a normal component. The obtained result declares that
some results on the dynamics of the cavity in superfluid helium should be reviewed
\end{abstract}

\maketitle

\section{Introduction and scientific background}

The study of the cavity dynamics is an important part of the problems of
continuum mechanics, including the hydrodynamics of superfluids. These may
be problems related to the evolution of bubbles created by electrons, (see,
e.g., \cite{Salomaa1981},\cite{Tempere2003},\cite{Guo2007}), bubbles caused
by sound (see, e.g., \cite{ABE2009}), cavitation due to negative pressure
(see, e.g.,\cite{Maris2000}), collapse of bubbles (see, e.g.,\cite{Qu2016},%
\cite{Qu2017}), sonoluminescence (see, e.g.,\cite{Jarman1966}), etc. Another
series of examples is related to the heat transfer in He II and to the
possibility of utilizing superfluid helium as a coolant in cryogenic
systems, which has been discussed extensively recently (see, e.g., book \cite%
{VanSciver2013}). Knowledge of the laws governing the formation and
development of vapor films on the surfaces of heaters is important for
solving corresponding problems ( see, e.g., \cite{Kryukov2006},\cite%
{Kryukov2013},\cite{Takada2014},\cite{Grunt2019}).

\qquad Studying the dynamics of the cavity, the authors of the above works
appeal to the Rayleigh-Plesset problem on the evolution and oscillations of
an air or vapor bubble, elaborated initially for an ordinary fluid (see,
e.g., book \cite{VENakoryakov1990} and references therein). Such treatment,
however, is justified in the case when superfluid helium behaves as an
ordinary fluid and moves as a whole with a mass velocity $\mathbf{v}=\mathbf{%
j}/\rho $ (see the notations below). This situation occurs when helium is
driven in the motion under the action of a pressure gradient (or gravity).
However, whenever the heat fluxes occur, a flow of the normal component
appears with a velocity\textbf{\ }$\mathbf{v}_{n}$, different from the mass
velocity $\mathbf{v}$ and the problem requires a fundamentally different
treatment related to the two-fluid nature of He II.

In the present paper the problem of evolution of a spherically symmetric
vapor bubble in the superfluid helium is considered. The equation playing
the role of the famous Rayleigh-Plesset problem in an ordinary fluid is
derived. The obtained equation results in a number of effects absent in
ordinary fluids. They include effects such as additional pressure caused by
the movement of the normal component or an extra-damping term. These effects
essentially influence the dynamics of cavities and should be taken into
account in the relevant works. \ To our knowledge, the analog of the
classical Rayleigh--Plesset equation for the two-fluid hydrodynamics of
superfluid helium have been not previously considered.

Loosely speaking the dynamics of bubble is determined by inertia (mass) of
ambient liquid and by the elastic properties due to pressure inside cavity.
The latter is result of many factors including such as surface tension,
Coulomb pressure, hydrostatic pressure, viscous contribution into stress
tensor etc. The corresponding task is a complex problem requiring careful
analysis of many factors.

In this work, we will concentrate on hydrodynamics processes inside the
fluid, not considering the involved phenomena inside the bubble. In
addition, for clarity and for deductive purposes, we take a pure two-fluid
Landau-Khalatnikov model, excluding more complex phenomena, such as quantum
turbulence (see, e.g., \cite{Nemirovskii2013},\cite{Nemirovskii2019a})or
existence of tiny thermal boundary layer near the interface boundary
liquid-vapor (see, e.g., \cite{Putterman1974}). Realizing the importance of
the above processes on the whole evolution of the cavity, we nonetheless
omit them in order to highlight the fascinated features of the two-fluid
Landau-Khalatnikov model of superfluid hydrodynamics in the process under
consideration.

The structure of the paper is as follows: The second and third sections can
be considered as the introductory. They are useful to introduce basic ideas
and methods, notations and terminology. In Sec. II we shortly describe the
Rayleigh problem in ordinary (nonsuperfluid)\ fluid and in Sec. III we write
down and comment a set of the motion equation of superfluid hydrodynamics.
Sec. IV is devoted to a detailed derivation of the Rayleigh--Plesset like
equation for the evolution of the boundary position of a spherical bubble in
superfluid helium. The main features of the derived equation are also
described there. In Conclusion the remarks on the obtained results are made
and possible development is discussed.

\section{The vapor film dynamic in ordinary (nonsuperfluid) fluids}

Before starting our study on the bubble dynamic\ in superfluid helium it
seems instructive to briefly describe the corresponding Rayleigh problem in
the ordinary (nonsuperfluid) fluid (see, e.g. book \cite{VENakoryakov1990}).
In spherical coordinates the continuity equation for velocity $v$ reads (the
fluid is supposed to be incompressible)
\begin{equation}
\frac{\partial }{\partial r}(r^{2}v)=0.  \label{contin}
\end{equation}%
The obvious solution of equation \ref{contin} is
\begin{equation}
v(r,t)=\frac{R^{2}}{r^{2}}\frac{dR}{dt}.  \label{v classiic}
\end{equation}%
Here $R=R(t)$ is the time dependent radius of the bubble. Solution \ref{v
classiic} is a consequence of that the rate of change of the bubble radius $%
dR/dt$ coincides with the velocity $v(r=R,t)$, which justifies Eq. (\ref{v
classiic}). Here we neglect the motion of the boundary arising from the mass
flux density at the interface liquid-vapor due to evaporation (or
condensation) \cite{Grunt2019}.

The momentum equation in spherical coordinates reads
\begin{equation}
\frac{\partial v}{\partial t}+v\frac{\partial v}{\partial r}=-\frac{1}{\rho }%
\frac{\partial p}{\partial r}  \label{Euler}
\end{equation}

Substituting the solution (\ref{v classiic}) into expression (\ref{Euler})
and integrating over $r$ in limits from from the bubble boundary $r=R(t)$ to
$\infty $ we have.

\begin{equation}
R\frac{d^{2}R}{dt^{2}}+\frac{3}{2}\left( \frac{dR}{dt}\right)
^{2}=(p(R)-p_{\infty })/\rho .  \label{Rayleigh cl}
\end{equation}%
Here $p_{\infty }$ is the pressure far from the bubble, often the
hydrostatic pressure $\rho gh$. Thus, as it was described above, the
evolution of the vapor bubble occurs due to inertia of fluids and elastic
properties of pressure on the bubble boundary. The latter appears from many
factors, viscosity, surface tension etc (see, e.g. book \cite%
{VENakoryakov1990}). When the fluid viscosity and the surface tension are
taken into account, equation (\ref{Rayleigh cl}) is converted into the
classical Rayleigh--Plesset equation
\begin{equation}
R\frac{d^{2}R}{dt^{2}}+\frac{3}{2}\left( \frac{dR}{dt}\right) ^{2}+4\frac{%
\nu }{R}\frac{dR}{dt}=(p(R)-p_{\infty }-\frac{2\gamma }{R})/\rho ,
\label{RPE}
\end{equation}%
where $\nu $ is the kinematic viscosity and $\gamma $ is the surface
tension. \ The Rayleigh--Plesset equation (\ref{RPE}) is intended to
describe the evolution of the bubble, to determine the size of the bubbles,
the oscillation frequency, etc.

The principle problem is to determine the pressure $p(R)$\ at the interface.
For that it is necessary to use some additional considerations. In the
simplest case, it can be assumed that an adiabatic process occurs inside the
bubble. However, in general case, the pressure should be extracted from
resolution of adjoint thermal problem. For instance in series of works on
the boiling helium (\cite{Dergunov2000},\cite{Kryukov2006},\cite{Grunt2019}%
), the vapor pressure $p(R)$ is determined from the Boltzmann kinetic
equation for problems of evaporation and condensation. Thus, "mechanical"
and "thermal" problems are tied already in classical fluids. As we will see
later, this statement is more actual in case of superfluid helium.

\section{Hydrodynamics of quantum fluids. Two-fluid model}

Before we consider the dynamics of the vapor bubble in the superfluid
helium, it is useful to recall shortly the hydrodynamics of superfluid
helium. From the viewpoint of hydrodynamics, He II can be viewed as a
mixture of two components. One of them, a superfluid liquid with density $%
\rho _{s}(p,T)$ ($p$ and $T$ are the pressure and the temperature,
correspondingly), moves with velocity $\mathbf{v}_{s}$ . The superfluid
component has no shear viscosity, it cannot be subjected to torsion $(\nabla
\times \mathbf{v}_{s}=0)$, and also cannot absorb and carry heat. Another
part with the density $\rho _{n}(p,T)$, moving with velocity $\mathbf{v}_{n}$%
, is the normal component, it behaves as usual classical viscous fluid. From
a deeper point of view, the flow of normal component is just the drift of
the thermal excitations (phonons and rotons), which appeared in the
background coherent state The motion of the two components is
thermodynamically reversible and consequently independent. The superfluid
component $\rho _{s}(T)$ appears at below $T_{\lambda }\approx 2.1768$ K at
saturated vapor pressure, growing with the decrease of the temperature and
reaching the full density $\rho =\rho _{s}(T)+\rho _{n}(T)$ at zero
temperature. The full flux of mass is just
\begin{equation}
\mathbf{j=}\rho _{s}\mathbf{v}_{s}+\rho _{n}\mathbf{v}_{n}  \label{mass flux}
\end{equation}

The equations of motion of such a liquid can be obtained on the basis of the
laws of conservation \cite{Landau1941} (see also \cite{Khalatnikov1956},
\cite{Putterman1974}). We shall write out and explain these equations, here
we restrict ourselves with the dissipationless case:

\bigskip
\begin{equation}
\frac{\mathcal{\partial }{\rho }}{\mathcal{\partial }t}+\;{\nabla \cdot \ }%
\mathbf{j}~=~0\;,  \label{rho diss}
\end{equation}%
\begin{equation}
\frac{\mathcal{\partial }j_{i}}{\mathcal{\partial }t}\;+\;\frac{\mathcal{%
\partial }\Pi {_{ik}}}{\mathcal{\partial }{x_{k}}}=0\;,  \label{j diss}
\end{equation}%
\begin{equation}
\frac{\mathcal{\partial }{S}}{\mathcal{\partial }t}+\;{\nabla }\cdot \left[ S%
\mathbf{v}_{n}\,\right] ~=0,  \label{s diss}
\end{equation}%
\begin{equation}
\frac{\mathcal{\partial }\mathbf{v}_{s}}{\mathcal{\partial }t}+\,\,\nabla
(\mu +\frac{\mathbf{v}_{s}^{2}}{2})=0.  \label{vs diss}
\end{equation}%
Equations (\ref{rho diss}), (\ref{j diss}) are the usual laws of
conservation of mass and momentum density.

The momentum density consist of two part - superfluid and normal
\begin{equation}
j_{i}=\rho _{s}v_{si}+\rho _{n}v_{ni}.  \label{momentum}
\end{equation}%
\bigskip

The momentum flux tensor ${\Pi _{ik}}$ has also two ingredients and is equal
to%
\begin{equation}
{\Pi _{ik}=}\rho _{s}v_{si}v_{sk}+\rho _{n}v_{ni}v_{nk}+\delta _{ik}p
\label{flux tensor}
\end{equation}

The subscripts $i,k$ denote the coordinates $x,y,z$; \ $\delta _{ik}$is the
unit tensor. As can' be seen from (\ref{flux tensor}), the complete
momentum-flux tensor can be decomposed into a normal part and a superfluid
part, and Eq. (\ref{j diss}) has an obvious structure. Equation (\ref{s diss}%
) is also obvious. This is the law of conservation of density of entropy $%
S(p,T)$. Here we see reflected the fact that entropy is carried over only by
the normal component. The expression (\ref{vs diss}) for the velocity of the
superfluid component is new, in contrast with the expression for an ordinary
liquid. It contains the information that the superfluid component cannot be
subjected to torsion because it has no shear viscosity. Therefore, $\nabla
\times {\ }\mathbf{v}_{s}$, and consequently, the convection term is $({\ }%
\mathbf{v}_{s}\cdot \nabla ){\ }\mathbf{v}_{s}\,$ $=\nabla (\mathbf{v}%
_{s}^{2}/2)$. The driving force for the superfluid part is the chemical
potential $\mu (p,T)$.

We write an expression for the energy flux $\mathbf{W}$ which follows from
Eqs. (\ref{rho diss})-(\ref{vs diss}) and which will be useful in our
further discussion:%
\begin{equation}
\mathbf{W=}(\mu +\frac{\mathbf{v}_{s}^{2}}{2})\mathbf{j+}ST\mathbf{v}%
_{n}+\rho _{n}\mathbf{v}_{n}\left( (\mathbf{v}_{n}-\mathbf{v}_{s})\cdot
\mathbf{v}_{s}\right) +\mathbf{W}_{irr}  \label{Energy flux}
\end{equation}

Here $\mathbf{W}_{irr}$ stands for the irreversible fluxes caused by the
dissipative effects, which are negligibly small for all real cases. It
should be noted at once that we observe a macroscopic energy flux $ST\mathbf{%
v}_{n}$ even in the case when the total mass flow $\mathbf{j}$ is equal to
zero (the so called counterflow). Neglecting the nonlinear effects of third
order and taking that the energy flux\textsl{\ }$\mathbf{W}$\textsl{\ }in
this case is just the heat flux $\mathbf{Q}$ , we arrive at formula
\begin{equation}
\mathbf{Q=}ST\mathbf{v}_{n}  \label{heat flux}
\end{equation}

In case $v_{n}=v$ (and in accordance with Eq. (\ref{momentum}))\textbf{\ }$%
\mathbf{v}_{n}=\mathbf{v}_{s}=\mathbf{v}$\textbf{,} that is, the fluid moves
as a whole (the so called co-flow) the energy flux\textsl{\ }$\mathbf{W}$
has a form
\begin{equation}
\mathbf{W}=(\mu +\frac{\mathbf{v}_{s}^{2}}{2})\rho \mathbf{v}+ST\mathbf{v}%
=(\mu +\frac{\mathbf{v}_{s}^{2}}{2})\mathbf{v}+ST\mathbf{v}=(h+\frac{\mathbf{%
v}^{2}}{2})\mathbf{v.}  \label{W coflow}
\end{equation}%
(here $h=\mu +TS$ is the enthalpy) as it should be in the one-phase fluid or
in case of co-flow (see, e.g. the handbook \cite{Landau1987}).

\section{ \ Rayleigh--Plesset problem\ in superfluid helium.}

\subsection{Treatment of the momentum flux tensor $\Pi _{rr\text{ }}$}

From the solution of the problem in classical fluid it is seen that the
master variable, which controls the whole process is the mass flow velocity $%
\mathbf{v}(\mathbf{r},t)$, for the simple reason that in spherical case the
variable $v(r,t)$ at the boundary points coincides with quantity $dR/dt$.
The master Rayleigh--Plesset equation is derived from equation for momentum
density, which, due to non compressibility condition is just the Euler
equation (\ref{Euler}) for velocity $v(r,t)$.

In the superfluid helium situation is more complicated. The reason is that
the equation (\ref{j diss}) for momentum density $j_{i}$ (\ref{momentum}),
does not include directly the mass flow velocity $v(r,t)$ as it takes place
in case of ordinary fluids (see Eq. (\ref{Euler})). In other words the
following inequity takes place

\begin{equation}
\frac{\partial }{\partial r_{k}}\Pi _{ik}\neq \rho v_{k}\frac{\partial v_{i}%
}{\partial r_{k}},  \label{flux nonequility}
\end{equation}%
and as it is frequently used or understood in the relevant works. On the
contrary, because of the two fluid hydrodynamics, the structure of the
momentum flux density tensor is more involved and should reflect the
presence of two ingredients - the superfluid and normal parts.

In fact, the superfluid and normal velocities $\mathbf{v}_{s}$, $\mathbf{v}%
_{n}$ are not convenient for solving our problems. More suitable are
variables the mass flow velocity $\mathbf{v}(r,t)=\mathbf{j}/\rho =(\rho _{s}%
\mathbf{v}_{s}+\rho _{n}\mathbf{v}_{n})/\rho $ \ and the velocity $\mathbf{v}%
_{n}$ of normal component. Indeed, the mass flow velocity \ $\mathbf{v}(r,t)$
is responsible for inertia of fluid and the normal velocity $\mathbf{v}_{n}$%
\ is tightly related to thermal processes in accordance to relation (\ref%
{heat flux}). Further we take that the total density $\rho $, as well as the
superfluid and normal densities $\rho _{s}$ and $\rho _{n}$ separately, are
constant.

The next, crucial step is to treat equation for momentum density (\ref{j
diss}). We have to express the momentum flux tensor $\Pi _{ik\text{ }}$via
quantities $\mathbf{v}(r,t)$ and $\mathbf{v}_{n}(r,t)$, which were selected
as the primary variables. To\ get rid of the supefluids velocity entering in
equation (\ref{flux tensor}) for $\Pi _{ik\text{ }}$we use relation, known
from classical superfluid hydrodynamics (see books \cite{Khalatnikov1956},
\cite{Putterman1974}, and also \ Eq. (\ref{momentum})).

\begin{equation}
\mathbf{v}_{s}=\frac{\rho \mathbf{v}}{\rho _{s}}\mathbf{-}\frac{\rho _{n}}{%
\rho _{s}}\mathbf{v}_{n}  \label{vs via vn}
\end{equation}

The momentum flux tensor $\Pi _{ik\text{ }}$ (see Eq. (\ref{flux tensor}))
can be rewritten in the chosen variables $\mathbf{v}(r,t)$ and\textbf{\ }$%
v_{n}(r,t)$ as%
\begin{gather}
\Pi _{ik\text{ }}=\rho _{s}(\frac{\rho v_{i}}{\rho _{s}}\mathbf{-}\frac{\rho
_{n}}{\rho _{s}}v_{ni})(\frac{\rho v_{k}}{\rho _{s}}\mathbf{-}\frac{\rho _{n}%
}{\rho _{s}}v_{n_{k}})+\rho _{n}v_{ni}v_{nk}+\delta _{ik}p
\label{flux tensor via vn j} \\
=\frac{\rho ^{2}}{\rho _{s}}v_{i}v_{k}+\frac{\rho _{n}^{2}}{\rho _{s}}%
v_{nk}v_{ni}-\frac{\rho \rho _{n}}{\rho _{s}}v_{i}v_{nk}-\frac{\rho \rho _{n}%
}{\rho _{s}}v_{k}v_{in}+\rho _{n}v_{in}v_{kn}+p\delta _{ik}  \notag
\end{gather}

Now we have to transform the momentum flux density tensor\textsl{\ }$\Pi _{ik%
\text{ }}$(\ref{flux tensor via vn j})\textsl{\ }into spherical coordinates.
The simplest way to do this is as follows. We can represent expressions of
type $A_{k}\mathcal{\partial }B_{i}{/}\mathcal{\partial }{x_{k}}$ as a $i$
component of combination $(\mathbf{A}\nabla )\mathbf{B}$. In accordance with
well known mathematical relation $(\mathbf{A\nabla })\mathbf{B}=\mathbf{%
\nabla }(\mathbf{AB/}2\mathbf{)}$ (see, e.g., \cite{Korn1968})\textbf{.}
The latter operation is possible, since due to spherical symmetry and
incompressibility of both components, $\nabla \cdot \mathbf{A=}0$ , $\nabla
\times \mathbf{A}=0$ and the same for vector $\mathbf{B}$. Using this rule
we rewrite the radial component of the momentum flux density tensor\textsl{\
}$\Pi _{rr\text{ }}$(\ref{flux tensor via vn j})\textsl{\ }in spherical
coordinates as
\begin{equation}
\Pi _{rr\text{ }}(r)=\frac{1}{2\rho _{s}}\left( v^{2}\rho ^{2}-2\rho \rho
_{n}vv_{n}+\rho _{n}\rho v_{n}^{2}\right)  \label{flux tensor spherical}
\end{equation}

Then, the equation for momentum $j(r,t)$, or, better for the mass velocity $%
v(r,t)=$ $j(r,t)/\rho $, has the form:
\begin{equation}
\frac{\partial v}{\partial t}+\frac{1}{\rho }\frac{\partial p}{\partial r}+%
\frac{\rho }{\rho _{s}}v\frac{\partial v}{\partial r}-\frac{\rho _{n}}{\rho
_{s}}(v_{n}\frac{\partial v}{\partial r}+v\frac{\partial v_{n}}{\partial r})+%
\frac{\rho _{n}}{\rho _{s}}v_{n}\frac{\partial v_{n}}{\partial r}=0
\label{dv/dt via v vn}
\end{equation}

If $v_{n}=v$ (the co-flow case) the momentum flux density tensor\textsl{\ }$%
\Pi _{ik\text{ }}$(\ref{flux tensor spherical})\textsl{\ }transforms to

\begin{equation}
\Pi _{rr\text{ }}=\frac{1}{2}v^{2}\frac{\rho }{\rho _{s}}\left( \rho -\rho
_{n}\right) =\frac{1}{2}\rho v^{2}  \label{flux tensor via v vn}
\end{equation}%
as it should be in the ordinary fluid.

\subsection{\protect\bigskip\ Treatment of $\mathbf{v}_{n}$, Flux of energy}

Of course, the direct resolution of a set of equations (\ref{rho diss})-(\ref%
{vs diss}) with substitutions (\ref{vs via vn}) and (\ref{flux tensor
spherical}) is the most general way to study the problem of the cavity
dynamics in the superfluid helium. There is, however, the simpler approach
which allows to get rid of the normal velocity $\mathbf{v}_{n}$\ with the
use of the expression for the energy flux $\mathbf{W}$ (\ref{Energy flux}).
Variant of this way was used in paper \cite{Grunt2019} where the authors
used relation (\ref{heat flux}) to express $\mathbf{v}_{n}$ via heat flux $Q$%
, where $Q$ is energy released by heater. Moreover, referring to the Gorter
-Mellink regime they added the mass flow velocity $\mathbf{v}$ to quantity $%
\mathbf{v}_{n}$ (see Eq. (33) in paper \cite{Grunt2019} ). Since, however,
the total mass flow $\mathbf{j}$ is not equal to zero, part of total energy
released by heater is converted into mechanical energy, which associated
with motion of helium as a whole. Thus. the flow is not pure counterflow
(and not co-flow, either), the normal velocity $\mathbf{v}_{n}$ is not
determined unambiguously by the energy flux and and the situation requires
more thorough investigation. We have to use the full expression for the
energy flux (\ref{Energy flux}). Neglecting again the nonlinear effects of
third order and irreversible heat fluxes $\mathbf{W}_{irr}$ \textsl{\ }we
arrive at formula (since we work for pure spherically symmetric case we omit
the vector notations)

\begin{equation}
v_{n}=\frac{W}{ST}-\frac{\mu \rho }{ST}v  \label{vn via Q mu}
\end{equation}

The origin and nature of the energy flow can be different, for example, it
can be a spherical heater making up a (spherical) vapor region around itself
or it can be some additional external pressure that causes the vortex cavity
to either collapse or oscillate.

Substituting $v_{n}$ from \ref{vn via Q mu} in expression for the momentum
flux tensor ${\Pi _{rr}}$ (\ref{flux tensor spherical}) in spherical case we
obtain%
\begin{eqnarray}
\Pi _{rr\text{ }} &=&\frac{1}{\rho }\frac{1}{2\rho _{s}}(v^{2}\rho
^{2}-2v\rho \rho _{n}(\frac{W}{ST}-\frac{\mu \rho }{ST}v)+\rho _{n}\rho (%
\frac{W}{ST}-\frac{\mu \rho }{ST}v)^{2})  \label{P via mu} \\
&=&\allowbreak \frac{1}{2S^{2}T^{2}\rho _{s}}\left( S^{2}T^{2}v^{2}\rho
+2\rho _{n}STv^{2}\mu \rho -2\rho _{n}STvW+\rho _{n}v^{2}\mu ^{2}\rho
^{2}-2\rho _{n}v\mu \rho W+\rho _{n}W^{2}\right)  \notag
\end{eqnarray}

Further we will use instead of the chemical potential $\mu $\ the enthalpy $%
h=\mu \rho +ST.$The enthalpy $h$ is more reliably measured and tabulated.
After that the expression for $\Pi _{rr\text{ }}$ take a form.

\begin{gather}
\Pi _{rr\text{ }}=\left( \frac{1}{2S^{2}T^{2}\rho _{s}}\left( h^{2}\rho
_{n}+S^{2}T^{2}\rho -S^{2}T^{2}\rho _{n}\right) \right) v^{2}+
\label{P final} \\
-\left( \frac{\rho _{n}h}{S^{2}T^{2}\rho _{s}}\right) Wv\ +\frac{\rho _{n}}{%
2S^{2}T^{2}\rho _{s}}W^{2}  \notag
\end{gather}%
\bigskip

We grouped the terms as follows: the first term contains the squared
velocity $v$; the second term contains the cross term $Wv\ $; and, finally,
the third term contains an expression that does not contain the mass
velocity $v$ at all. coefficients Then, substituting (\ref{P final}) into
equation (\ref{dv/dt via v vn}) we get

\begin{equation}
\frac{\partial v}{\partial t}+\frac{1}{\rho }\frac{\partial p}{\partial r}+Av%
\frac{\partial v}{\partial r}+\frac{1}{2}Bv\frac{\partial W}{\partial r}+%
\frac{1}{2}BW\frac{\partial v}{\partial r}+C\frac{\partial }{\partial r}%
\frac{1}{2}W^{2}=0.  \label{eq for v 2}
\end{equation}%
$A,B$ and $C$ are equal%
\begin{equation*}
A=\frac{1}{2S^{2}T^{2}\rho _{s}}\left( h^{2}\rho _{n}+S^{2}T^{2}\rho
-S^{2}T^{2}\rho _{n}\right)
\end{equation*}%
\begin{equation*}
B=-\frac{\rho _{n}}{S^{2}T^{2}\rho _{s}}h
\end{equation*}%
\begin{equation*}
C=\frac{1}{S^{2}T^{2}}\frac{\rho _{n}}{\rho _{s}}
\end{equation*}%
The physical meaning of such a grouping we will discuss further

\subsection{\protect\bigskip Equation for the boundary position $R(t)$}

We proceed to derive equations for the boundary position evolution $R(t)$.
Just as it had been done in the case of classical fluid, we start with the
continuity equation (\ref{rho diss}). It includes only quantity $j(r,t)$ \
and it has the obvious solution,\ exactly as in ordinary classical fluids.

\begin{equation}
{\ j(r,t)=\rho }\frac{R^{2}}{r^{2}}\frac{dR}{dt}  \label{flux j}
\end{equation}

Again, due to incompressibility condition, the mass flow $v$\ velocity is $%
v=j/\rho $ and coincides with the classical solution \ref{v classiic}.

To move further we have, just as in the classical case, to work with the
equation for the momentum flux $\ j(r,t)$. For concretization, we consider a
purely thermal problem: the development of a vapor film created by a
spherical heater of the radius $R_{H}$, immersed in superfluid helium. Then,
the only source of energy is the heat released on the heater, and the energy
flux $W(r)$\ into the surrounding space has the form\textsl{\ }%
\begin{equation}
W=\frac{R_{H}^{2}}{r^{2}}Q.  \label{W(r)}
\end{equation}%
Here $Q$ is the heat flux density, released on the surface of the heater.
Combining (\ref{W(r)}) with the expression for mass velocity $v(r,t)=\frac{%
R^{2}}{r^{2}}\frac{dR}{dt}$ following from (\ref{flux j}) and accomplishing
differentiation with respect to $r$ we rewrite (\ref{eq for v 2})

\begin{equation*}
\frac{R^{2}}{r^{2}}\frac{d^{2}R}{dt^{2}}+\frac{2R}{r^{2}}(\frac{dR}{dt})^{2}+%
\frac{1}{\rho }\frac{\partial p}{\partial r}+A(-2)\frac{R^{2}}{r^{2}}\frac{dR%
}{dt}(\frac{R^{2}}{r^{3}}\frac{dR}{dt})+B((\frac{R_{H}^{2}}{r^{2}}Q)(-2)%
\frac{R^{2}}{r^{3}}\frac{dR}{dt}+\frac{R^{2}}{r^{2}}\frac{dR}{dt}(-2)\frac{%
R_{H}^{2}}{r^{3}}Q)+C\frac{R_{H}^{2}}{r^{2}}Q(-2)\frac{R_{H}^{2}}{r^{3}}Q=0.
\end{equation*}

Integrating this equation over $r$ in limits from $R$ to $\infty $ we have.
\
\begin{equation}
R\frac{d^{2}R}{dt^{2}}+(2-\frac{A}{2})\left( \frac{dR}{dt}\right) ^{2}-B(%
\frac{R_{H}^{2}}{R^{2}}Q)\frac{dR}{dt}-C(\frac{R_{H}^{2}}{R^{2}}Q)^{2}\frac{1%
}{2}+\frac{1}{\rho }(p(\infty )-p(R))\ =0  \label{after integral}
\end{equation}

Just as in case of a classical fluid, it is possible to add terms into
stress tensor (pressure) arising due to viscosity of the normal component
and surface tension. \ So the final form is:

\begin{equation}
R\frac{d^{2}R}{dt^{2}}+(2-\frac{A}{2})(\frac{dR}{dt})^{2}=-4\frac{\nu _{n}}{R%
}\frac{dR}{dt}+B(\frac{R_{H}^{2}}{R^{2}}Q)\frac{dR}{dt}+\frac{1}{2}C(\frac{%
R_{H}^{2}}{R^{2}}Q)^{2}+\frac{1}{\rho }(p(R)-p(\infty )-\frac{2\gamma }{R})
\label{RP final}
\end{equation}%
Here $\nu _{n}$ is the shear viscosity of the normal component. The master
equation for the boundary position of the film (\ref{RP final}), plays the
role of Rayleigh--Plesset equation for superfluid helium. We again note that
some processes such as quantum turbulence (see, e.g., \cite{Nemirovskii2013}%
) or the thermal boundary layer (see, e.g., \cite{Putterman1974}) were not
included into consideration of the whole problem. In this sense, the
equation (\ref{RP final}) can be considered as the first step.

\subsection{Analysis of the Rayleigh--Plesset equation}

Equation (\ref{RP final}) differs from the Rayleigh--Plesset equation for
ordinary fluids \ref{RPE}. It includs additional terms, absent in the
classical case. This difference arises due to two fluid model and specific
form of the momentum flux tensor $\Pi _{rr\text{ }}$ (\ref{flux tensor}). In
the case when $v_{n}=v$ (and in accordance with Eq. (\ref{momentum}))\textbf{%
\ }$\mathbf{v}_{n}=\mathbf{v}_{s}=\mathbf{v}$\textbf{,} that is, the fluid
moves as a whole, the co-flow case) Eq. (\ref{RP final}) is reduced to the
classical equation \ref{RPE}.Indeed, if to use $W=hv$ (neglecting the
nonlinear effects of third order), then $\Pi _{rr\text{ }}$ (\ref{P final})
transforms as

\begin{equation*}
\frac{1}{2S^{2}T^{2}}\frac{v^{2}}{\rho _{s}}\left( S^{2}T^{2}\rho -2\rho
_{n}STh+2\rho _{n}ST(h-ST)+\rho _{n}h^{2}-2\rho _{n}h(h-ST)+\rho
_{n}(h-ST)^{2}\right) =\frac{1}{2}v^{2}\ ,
\end{equation*}%
that is, it acquires its classical value and equation (\ref{RP final}) is
converted into equation \ref{RPE}. As a result, the well-known problems such
as isothermal oscillations of the gas bubble, or the collapse of empty
cavity, fully coincides with the classical solution (see, e.g., book \cite%
{VENakoryakov1990}). The real and essential difference appear for non-zero
heat transfer.

The terms in the momentum flux tensor $\Pi _{rr\text{ }}$and, hence, in
equation (\ref{RP final}) are combined into groups with different physical
meanings. So the terms containing derivatives $dR/dt$ are important for
non-stationary processes, such as a transient process, or oscillatory
motion. The remaining terms not containing derivative $dR/dt$ determine the
stationary solution $R(t\rightarrow \infty )$, e.g. the thickness of the
vapor film (or size of the vapor bubble). So, the third term in the right
hand side of Eq. (\ref{RP final}), which is independtnt on the velocity of
the boundary position $dR/dt$ , can be unified with the pressure term $p(R)$%
. That can be additionally justified by the fact that this term is
proportional to squared velocities and, hence, it is related to dependence
of pressure on the velocity (see also books \cite{Khalatnikov1956}, \cite%
{Putterman1974}). Further we will name it as "Bernoulli" -like pressure. In
many cases, this additional pressure is small, for example, in experiments (%
\cite{Dergunov2000},\cite{Kryukov2006}) \textsl{\ }this "Bernoulli" -like
pressure is of the order of 10\% of the hydrostatic pressure $\rho gh$,
however for smaller values of $h$ , and, especially, under microgravity
conditions it can be extremely important.

The second term in the right hand side of Eq. (\ref{RP final}) is of the
particular interest. It has the same structure as the viscous term with
shear viscosity for normal velocity, which is responsible for the
attenuation of bubble oscillations. At the same time it essentially (by
several ordersof magnitude) exceeds the usual viscous damping. For this
reason we will name it as "extra-damping" term. The described term can be
the reason for the strong attenuation of the bubble oscillations, observed
in many works (see, e.g., (\cite{Dergunov2000},\cite{Kryukov2006}). The
authors called it as abnormal "Suppression of oscillations of the
vapor--liquid phase boundary in superfluid helium" (see \cite{Kryukov2013}).
To our knowledge, the authors could not explain this phenomena and referred
to pure experimental obstacles.

Finally the second term in the in the left hand side of Eq. (\ref{RP final})
differs from classical case in that the coefficient $3/2$ is changed with
the quantity $2-A/2$. Preliminary numerical analysis of the solution of Eq. (%
\ref{RP final}) shows that this term does not affect greatlly the final
results.

\section{Conclusion}

The problem of the cavity dynamics in the superfluid helium is considered on
the basis of Landau-Khalatnikov two-fluid \ hydrodynamics. The equation,
which play the role of the classical Rayleigh--Plesset equation (\ref{RP
final}), significantly differs from its classical analogue. This difference
appears from special treatment of the momentum flux tensor, which, due to
the two-fluid nature of superfluids generates several new effects, such as a
"Bernoulli" -like pressure term or an extra-damping term. These terms
essentially affect the dynamics of cavities compared to ordinary fluids and
can influence results and conlusions made in in the relevant works.

Equation (\ref{RP final}) is intended to investigate problems associated
with the evolution of a cavity in superfluids. We again would like to
emphasize that this hydrodynamic description is part of the general problem
and, maybe, not the primary part. Probably more important ingredient is the
correct analysis of the pressure drop, due to the involved thermal or/and
electric processes inside the bubble.

Thus, in works on the boiling helium (see, e.g., \cite{Dergunov2000},\cite%
{Kryukov2006}), the authors determine vapor pressure $p(R)$ from the
Boltzmann kinetic equation for evaporation and condensation
processes.Studying multielectron bubbles in liquid helium (see, e.g., \cite%
{Salomaa1981},\cite{Tempere2003},\cite{Guo2007}), the authors find the
electron density and its contribution into the pressure inside the bubble
with the use of the Poisson equation.

The study of according processes is a separate involved problem, that goes
beyond the scope of this work. Therefore, we deliberately limited ourselves
to the hydrodynamic part, since our goal was to emphasize the role of
two-fluid hydrodynamics. Moreover, consequently pursuing that goal, we
simplified situation by taking the pure two-fluid Landau-Khalatnikov model
and omitting other phenomena such as quantum turbulence \cite%
{Nemirovskii2013} or existence of tiny thermal boundary layer near the
interface liquid-vapor \cite{Putterman1974}. These topics are supposed to be
study in future.

I would like to express my gratitude to Professor A. P. Kryukov, who has
drew my attention to the abnormal damping of oscillations of the vapor film
in helium, and I would also like to thank the staff of the Department of Low
Temperatures of the National Research University MPEI, for the discussion of
the work.

This work was supported by the Russian Science Foundation (grant No.
19-19-00321).

%\bibliographystyle{apsrev}
%\bibliography{E:/AASERGEY/PAPERS/1BIBTEX/QT}

\begin{thebibliography}{22}
\expandafter\ifx\csname natexlab\endcsname\relax\def\natexlab#1{#1}\fi
\expandafter\ifx\csname bibnamefont\endcsname\relax
  \def\bibnamefont#1{#1}\fi
\expandafter\ifx\csname bibfnamefont\endcsname\relax
  \def\bibfnamefont#1{#1}\fi
\expandafter\ifx\csname citenamefont\endcsname\relax
  \def\citenamefont#1{#1}\fi
\expandafter\ifx\csname url\endcsname\relax
  \def\url#1{\texttt{#1}}\fi
\expandafter\ifx\csname urlprefix\endcsname\relax\def\urlprefix{URL }\fi
\providecommand{\bibinfo}[2]{#2}
\providecommand{\eprint}[2][]{\url{#2}}

\bibitem[{\citenamefont{Salomaa and Williams}(1981)}]{Salomaa1981}
\bibinfo{author}{\bibfnamefont{M.~M.} \bibnamefont{Salomaa}} \bibnamefont{and}
  \bibinfo{author}{\bibfnamefont{G.~A.} \bibnamefont{Williams}},
  \bibinfo{journal}{Phys. Rev. Lett.} \textbf{\bibinfo{volume}{47}},
  \bibinfo{pages}{1730} (\bibinfo{year}{1981}),
  \urlprefix\url{https://link.aps.org/doi/10.1103/PhysRevLett.47.1730}.

\bibitem[{\citenamefont{Tempere et~al.}(2003)\citenamefont{Tempere, Silvera,
  and Devreese}}]{Tempere2003}
\bibinfo{author}{\bibfnamefont{J.}~\bibnamefont{Tempere}},
  \bibinfo{author}{\bibfnamefont{I.~F.} \bibnamefont{Silvera}},
  \bibnamefont{and} \bibinfo{author}{\bibfnamefont{J.~T.}
  \bibnamefont{Devreese}}, \bibinfo{journal}{Phys. Rev. B}
  \textbf{\bibinfo{volume}{67}}, \bibinfo{pages}{035402}
  (\bibinfo{year}{2003}),
  \urlprefix\url{https://link.aps.org/doi/10.1103/PhysRevB.67.035402}.

\bibitem[{\citenamefont{Guo et~al.}(2007)\citenamefont{Guo, Jin, and
  Maris}}]{Guo2007}
\bibinfo{author}{\bibfnamefont{W.}~\bibnamefont{Guo}},
  \bibinfo{author}{\bibfnamefont{D.}~\bibnamefont{Jin}}, \bibnamefont{and}
  \bibinfo{author}{\bibfnamefont{H.~J.} \bibnamefont{Maris}},
  \bibinfo{journal}{Journal of Physics: Conference Series}
  \textbf{\bibinfo{volume}{92}}, \bibinfo{pages}{012001}
  (\bibinfo{year}{2007}),
  \urlprefix\url{https://doi.org/10.1088%2F1742-6596%2F92%2F1%2F012001}.

\bibitem[{\citenamefont{ABE et~al.}(2009)\citenamefont{ABE, MORIKAWA, UEDA,
  NOMURA, OKUDA, and BURMISTROV}}]{ABE2009}
\bibinfo{author}{\bibfnamefont{H.}~\bibnamefont{ABE}},
  \bibinfo{author}{\bibfnamefont{M.}~\bibnamefont{MORIKAWA}},
  \bibinfo{author}{\bibfnamefont{T.}~\bibnamefont{UEDA}},
  \bibinfo{author}{\bibfnamefont{R.}~\bibnamefont{NOMURA}},
  \bibinfo{author}{\bibfnamefont{Y.}~\bibnamefont{OKUDA}}, \bibnamefont{and}
  \bibinfo{author}{\bibfnamefont{S.~N.} \bibnamefont{BURMISTROV}},
  \bibinfo{journal}{Journal of Fluid Mechanics} \textbf{\bibinfo{volume}{619}},
  \bibinfo{pages}{261–275} (\bibinfo{year}{2009}).

\bibitem[{\citenamefont{Maris and Balibar}(2000)}]{Maris2000}
\bibinfo{author}{\bibfnamefont{H.}~\bibnamefont{Maris}} \bibnamefont{and}
  \bibinfo{author}{\bibfnamefont{S.}~\bibnamefont{Balibar}},
  \bibinfo{journal}{Physics Today} \textbf{\bibinfo{volume}{53}},
  \bibinfo{pages}{29} (\bibinfo{year}{2000}).

\bibitem[{\citenamefont{Qu et~al.}(2016)\citenamefont{Qu, Trimeche, Jacquier,
  and Grucker}}]{Qu2016}
\bibinfo{author}{\bibfnamefont{A.}~\bibnamefont{Qu}},
  \bibinfo{author}{\bibfnamefont{A.}~\bibnamefont{Trimeche}},
  \bibinfo{author}{\bibfnamefont{P.}~\bibnamefont{Jacquier}}, \bibnamefont{and}
  \bibinfo{author}{\bibfnamefont{J.}~\bibnamefont{Grucker}},
  \bibinfo{journal}{Phys. Rev. B} \textbf{\bibinfo{volume}{93}},
  \bibinfo{pages}{174521} (\bibinfo{year}{2016}),
  \urlprefix\url{https://link.aps.org/doi/10.1103/PhysRevB.93.174521}.

\bibitem[{\citenamefont{Qu}(2017)}]{Qu2017}
\bibinfo{author}{\bibfnamefont{A.}~\bibnamefont{Qu}}, Ph.D. thesis
  (\bibinfo{year}{2017}).

\bibitem[{\citenamefont{Jarman and Taylor}(1966)}]{Jarman1966}
\bibinfo{author}{\bibfnamefont{P.~D.} \bibnamefont{Jarman}} \bibnamefont{and}
  \bibinfo{author}{\bibfnamefont{K.~J.} \bibnamefont{Taylor}},
  \bibinfo{journal}{The Journal of the Acoustical Society of America}
  \textbf{\bibinfo{volume}{39}}, \bibinfo{pages}{584} (\bibinfo{year}{1966}),
  \eprint{https://doi.org/10.1121/1.1909933},
  \urlprefix\url{https://doi.org/10.1121/1.1909933}.

\bibitem[{\citenamefont{Van~Sciver}(2013)}]{VanSciver2013}
\bibinfo{author}{\bibfnamefont{S.}~\bibnamefont{Van~Sciver}},
  \emph{\bibinfo{title}{Helium Cryogenics}}, International Cryogenics Monograph
  Series (\bibinfo{publisher}{Springer US}, \bibinfo{year}{2013}), ISBN
  \bibinfo{isbn}{9781489904997}.

\bibitem[{\citenamefont{Kryukov and Mednikov}(2006)}]{Kryukov2006}
\bibinfo{author}{\bibfnamefont{A.~P.} \bibnamefont{Kryukov}} \bibnamefont{and}
  \bibinfo{author}{\bibfnamefont{A.~F.} \bibnamefont{Mednikov}},
  \bibinfo{journal}{Journal of Applied Mechanics and Technical Physics}
  \textbf{\bibinfo{volume}{47}}, \bibinfo{pages}{836} (\bibinfo{year}{2006}),
  ISSN \bibinfo{issn}{1573-8620},
  \urlprefix\url{https://doi.org/10.1007/s10808-006-0122-0}.

\bibitem[{\citenamefont{Kryukov and Puzina}(2013)}]{Kryukov2013}
\bibinfo{author}{\bibfnamefont{A.}~\bibnamefont{Kryukov}} \bibnamefont{and}
  \bibinfo{author}{\bibfnamefont{Y.}~\bibnamefont{Puzina}}, \bibinfo{journal}{J
  Eng Phys Thermophy} \textbf{\bibinfo{volume}{86}} (\bibinfo{year}{2013}).

\bibitem[{\citenamefont{Takada et~al.}(2014)\citenamefont{Takada, Kimura,
  Mamiya, Okamura, Nozawa, and Murakami}}]{Takada2014}
\bibinfo{author}{\bibfnamefont{S.}~\bibnamefont{Takada}},
  \bibinfo{author}{\bibfnamefont{N.}~\bibnamefont{Kimura}},
  \bibinfo{author}{\bibfnamefont{M.}~\bibnamefont{Mamiya}},
  \bibinfo{author}{\bibfnamefont{T.}~\bibnamefont{Okamura}},
  \bibinfo{author}{\bibfnamefont{M.}~\bibnamefont{Nozawa}}, \bibnamefont{and}
  \bibinfo{author}{\bibfnamefont{M.}~\bibnamefont{Murakami}},
  \bibinfo{journal}{AIP Conference Proceedings}
  \textbf{\bibinfo{volume}{1573}}, \bibinfo{pages}{292} (\bibinfo{year}{2014}),
  \eprint{https://aip.scitation.org/doi/pdf/10.1063/1.4860714},
  \urlprefix\url{https://aip.scitation.org/doi/abs/10.1063/1.4860714}.

\bibitem[{\citenamefont{Grunt et~al.}(2019)\citenamefont{Grunt, Lewkowicz,
  Pietrowicz, Takada, Kimura, and Murakami}}]{Grunt2019}
\bibinfo{author}{\bibfnamefont{K.}~\bibnamefont{Grunt}},
  \bibinfo{author}{\bibfnamefont{M.}~\bibnamefont{Lewkowicz}},
  \bibinfo{author}{\bibfnamefont{S.}~\bibnamefont{Pietrowicz}},
  \bibinfo{author}{\bibfnamefont{S.}~\bibnamefont{Takada}},
  \bibinfo{author}{\bibfnamefont{N.}~\bibnamefont{Kimura}}, \bibnamefont{and}
  \bibinfo{author}{\bibfnamefont{M.}~\bibnamefont{Murakami}},
  \bibinfo{journal}{International Journal of Heat and Mass Transfer}
  \textbf{\bibinfo{volume}{134}}, \bibinfo{pages}{1073 }
  (\bibinfo{year}{2019}), ISSN \bibinfo{issn}{0017-9310},
  \urlprefix\url{http://www.sciencedirect.com/science/article/pii/S0017931018353328}.

\bibitem[{\citenamefont{VE~Nakoryakov}(1990)}]{VENakoryakov1990}
\bibinfo{author}{\bibfnamefont{I.~S.} \bibnamefont{VE~Nakoryakov},
  \bibfnamefont{BG~Pokusaev}}, \emph{\bibinfo{title}{Wave dynamics of gas-and
  vapor-liquid media}} (\bibinfo{publisher}{Energoizdat, Moscow,},
  \bibinfo{year}{1990}).

\bibitem[{\citenamefont{Nemirovskii}(2013)}]{Nemirovskii2013}
\bibinfo{author}{\bibfnamefont{S.~K.} \bibnamefont{Nemirovskii}},
  \bibinfo{journal}{Physics Reports} \textbf{\bibinfo{volume}{524}},
  \bibinfo{pages}{85 } (\bibinfo{year}{2013}).

\bibitem[{\citenamefont{Nemirovskii}(2019)}]{Nemirovskii2019a}
\bibinfo{author}{\bibfnamefont{S.~K.} \bibnamefont{Nemirovskii}},
  \bibinfo{journal}{Low Temperature Physics} \textbf{\bibinfo{volume}{45}},
  \bibinfo{pages}{841} (\bibinfo{year}{2019}),
  \eprint{https://doi.org/10.1063/1.5116532},
  \urlprefix\url{https://doi.org/10.1063/1.5116532}.

\bibitem[{\citenamefont{Putterman}(1974)}]{Putterman1974}
\bibinfo{author}{\bibfnamefont{S.}~\bibnamefont{Putterman}},
  \emph{\bibinfo{title}{Superfluid Hydrodynamics}}
  (\bibinfo{publisher}{North-Holland, Amsterdam}, \bibinfo{year}{1974}).

\bibitem[{\citenamefont{Dergunov et~al.}(2000)\citenamefont{Dergunov, Kryukov,
  and Gorbunov}}]{Dergunov2000}
\bibinfo{author}{\bibfnamefont{I.}~\bibnamefont{Dergunov}},
  \bibinfo{author}{\bibfnamefont{A.}~\bibnamefont{Kryukov}}, \bibnamefont{and}
  \bibinfo{author}{\bibfnamefont{A.}~\bibnamefont{Gorbunov}},
  \bibinfo{journal}{Journal of Low Temperature Physics}
  \textbf{\bibinfo{volume}{119}}, \bibinfo{pages}{403} (\bibinfo{year}{2000}).

\bibitem[{\citenamefont{Landau}(1941)}]{Landau1941}
\bibinfo{author}{\bibfnamefont{L.~D.} \bibnamefont{Landau}},
  \bibinfo{journal}{J. Phys.} \textbf{\bibinfo{volume}{5}}, \bibinfo{pages}{71}
  (\bibinfo{year}{1941}).

\bibitem[{\citenamefont{Khalatnikov}(1956)}]{Khalatnikov1956}
\bibinfo{author}{\bibfnamefont{I.~M.} \bibnamefont{Khalatnikov}},
  \bibinfo{journal}{Sov. Phys. JETP} \textbf{\bibinfo{volume}{3}},
  \bibinfo{pages}{649} (\bibinfo{year}{1956}).

\bibitem[{\citenamefont{Landau and Lifshitz}(1987)}]{Landau1987}
\bibinfo{author}{\bibfnamefont{L.}~\bibnamefont{Landau}} \bibnamefont{and}
  \bibinfo{author}{\bibfnamefont{E.}~\bibnamefont{Lifshitz}},
  \emph{\bibinfo{title}{Vol. 6. Fluid Mechanics, 2nd edition.}}
  (\bibinfo{publisher}{Pergamon Press,Oxford}, \bibinfo{year}{1987}),
  \bibinfo{edition}{third edition} ed.

\bibitem[{\citenamefont{Korn and Korn}(1968)}]{Korn1968}
\bibinfo{author}{\bibfnamefont{G.~A.} \bibnamefont{Korn}} \bibnamefont{and}
  \bibinfo{author}{\bibfnamefont{T.~M.} \bibnamefont{Korn}},
  \emph{\bibinfo{title}{Mathematical Handbook for Scientists and Engineers.}}
  (\bibinfo{publisher}{New York/San Francisco/Toronto/London/Sydney},
  \bibinfo{year}{1968}).

\end{thebibliography}

\end{document}